\documentclass[aps,showpacs,preprintnumbers,amsmath,amssymb�eqsecnum,
twocolumn, tightenlines,
]{revtex4}

\usepackage[dvips]{graphicx}
\usepackage{bm}
\usepackage{longtable}
\usepackage{amsmath}
\usepackage{dcolumn}
\usepackage{placeins}

\begin{document}

\bibliographystyle{apsrev}

\title {Screening and finite size corrections to the octupole and Schiff   moments}

\author{V. V. Flambaum} \email[Email:]{flambaum@phys.unsw.edu.au}
\author{A. Kozlov} \email[Email:]{o.kozloff@student.unsw.edu.au}

\affiliation{School of Physics, University of New South Wales, Sydney
  2052, Australia}

    \date{\today}

    \begin{abstract}
Parity (P) and time reversal (T) violating nuclear forces create $P,T$-odd moments in expansion of the nuclear electrostatic potential. We derive expression for the nuclear electric octupole field which includes the electron screening correction (similar to the screening term in the Schiff moment). Then we calculate  the $Z^2 \alpha^2$ corrections to the Schiff moment which appear due to the finite nuclear size. Such corrections are  important in heavy atoms with nuclear charge $Z>50$. 
The Schiff and octupole moments induce atomic electric dipole moments (EDM) and $P,T$-odd
interactions in molecules which are measured in numerous  experiments to test  $CP$-violation theories.
    \end{abstract}


\maketitle

 The measurements of EDM induced by parity- and time-violating forces provide crucial tests of modern unification theories.  The best limits on $P,T$-odd nuclear forces are obtained from the measurements of  $^{199}$Hg atomic EDM \cite{Romalis:2001}. Nuclear EDM can not induce atomic EDM  due to complete electron screening of nuclear EDM (Schiff theorem \cite{Purcell:1950, Schiff:1963, Sandars:1967}). The first $P,T$-odd terms that survive the screening and produce atomic EDM are the Schiff and octupole moments \cite{Sushkov:1984}. 
In this paper we  consider the partial screening of the nuclear electric octupole moment  and the finite nuclear size corrections to the Schiff moment.

  The nuclear electrostatic potential with electron screening taken into account can be written in the following form (see e.g. \cite{Auerbach:1997} for the derivation):
\begin{equation}\label{ScrPot}
\varphi ({\bf R})\!=\!Ze\left\{\int\!\frac{\rho({\bf r})}{|{\bf R}-{\bf r}|}d^3r + \langle{\bf r\rangle\cdot\nabla}\!\int\!\frac{\rho({\bf r})}{|{\bf R}-{\bf r}|}d^3r\right\},
\end{equation}
where $\int\rho({\bf r})d^3r=1$,  and ${\bf d}=Ze\langle{\bf r}\rangle=Ze\int\rho({\bf r}){\bf r}d^3r$ is the nuclear EDM. The second term cancels the dipole long-range electric field in the multiple expansion of $\varphi({\bf R})$.  We expand the Coulomb potential in the terms of Legendre polynomials
\begin{equation}
\frac{1}{|\bf R - r|}=\sum_l r_<^l/r_>^{l+1}P_l(\cos \theta),
\end{equation}
where $r_<$ and $r_>$ are min$\left[r, R\right]$ and max$\left[r, R\right]$ respectively,  $\theta$ is the angle between vectors ${\bf r}$ and ${\bf R}$. The Legendre polynomials can be written in the following form: $P_1(\cos\theta)=r_iR_i/(rR)$, $P_2(\cos\theta)=R_iR_j\cdot q_{ij}/(2r^2R^2)$, where $q_{ij}=3 r_i r_j- r^2 \delta_{ij}$ is the  quadrupole moment tensor (summation over  repeating indexes is assumed). The $P,T$-odd part of the potential (\ref{ScrPot}) originates from the odd harmonics {\it l} of the first term and even harmonics of the second term. We start our consideration of $P,T$-odd part $\varphi^{(1)}({\bf R})$ of (\ref{ScrPot}) with $l=1$ in the first term and $l=0, 2$ in the second term.  The dipole part of $\varphi^{(1)}({\bf R})$ corresponds to the Schiff moment field. The third harmonic $l=3$ in the first term of (\ref{ScrPot}) gives the octupole field that has been considered in \cite{Flambaum:1997}. We will add $l=3$ later. As we will show below, account of $l=2$ in the second term of (\ref{ScrPot}) gives the screening of the octupole field. We can present the $P,T$-odd part of the potential (\ref{ScrPot}) in the form
\begin{multline}\label{OddPot}
\phi^{(1)}({\bf R})=Ze\left[\frac{R_i}{R^3}\int_0^R\rho({\bf r})r_id^3r+R_i\int_R^{\infty}\frac{r_i}{r^3}\rho({\bf r})d^3r\right.\\
-\langle r_i\rangle\frac{R_i}{R^3}\int_0^R\rho({\bf r})d^3r+\langle r_j\rangle R_i\int_R^{\infty}\frac{q_{ij}}{r^5}\rho({\bf r})d^3r\\
\left.+\left(\frac{\langle r_i\rangle R_j}{R^5}-\frac{5}{2}\frac{\langle r_k\rangle R_i R_j R_k}{R^7}\right)\int_0^R q_{ij}\rho({\bf r})d^3r\right]
\end{multline}
Note that for $R\rightarrow\infty$ the first and third terms of Eq. (\ref{OddPot}) cancel each other. Therefore, for their sum we can use $\int_0^R=\int_0^{\infty}-\int_R^{\infty}=-\int_R^{\infty}$ and present $\phi^{(1)}$ as 
\begin{multline}\label{OddPot1}
\phi^{(1)}=Ze\!\left[ R_i\!\int_R^{\infty}\!\!\left(\frac{\langle r_i\rangle}{R^3}-\frac{r_i}{R^3}+\frac{r_i}{r^3}+\frac{\langle r_j\rangle q_{ij}}{r^5}\right)\!\rho({\bf r})d^3r\right.\\
\left.+\left(\frac{\langle r_i\rangle R_j}{R^5}-\frac{5}{2}\frac{\langle r_k\rangle R_i R_j R_k}{R^7}\right)\int_0^R q_{ij}\rho({\bf r})d^3r\right].
\end{multline}
Last  term in the above equation can be presented as
\begin{multline}\label{tens}
\frac{\langle r_i\rangle R_j}{R^5}-\frac{5}{2}\frac{\langle r_k\rangle R_i R_j R_k}{R^7}=-\frac{5}{2R^7}O_{ijk}\langle r_k\rangle+\\
\left\{\frac{\langle r_i\rangle R_j}{R^5}-\frac{\langle r_k\rangle}{2R^5}(\delta_{ij}R_k+\delta_{jk}R_i+\delta_{ki}R_j)\right\},
\end{multline}
where
\begin{equation}
O_{ijk}=\left[R_i R_j R_k-\frac{R^2}{5}(\delta_{ij}R_k+\delta_{jk}R_i+\delta_{ki}R_j)\right].
\end{equation}
In Eq. (\ref{tens}) the  tensor in the brackets $\{...\}$ vanishes after convolution with the symmetric tensor $q_{ij}$ in Eq. (\ref{OddPot1}). Introducing $O_{ijk}$ into Eq. (\ref{OddPot1}) we obtain the equation for the $P,T$-odd part of the electrostatic nuclear potential
\begin{multline}\label{FinPot}
\phi^{(1)}=\!Ze R_i\!\int_R^{\infty}\!\!\left(\frac{\langle r_i\rangle}{R^3}-\frac{r_i}{R^3}+\frac{r_i}{r^3}+\frac{\langle r_j\rangle q_{ij}}{r^5}\right)\!\rho({\bf r})d^3r\\
-\frac{5}{2}\frac{Ze}{R^7}O_{ijk}\langle r_k \rangle\int_0^R q_{ij}\rho({\bf r})d^3r
\end{multline}
The last term in the above expression originates  from the second (screening) term in Eq. (\ref{ScrPot}). It gives the screening for the octupole field. The octupole appears due to the  third harmonic $l=3$ in the  Coulomb potential expansion and for $R>R_N$ is given by the equation:
\begin{equation}
\varphi^{(octupole)}({\bf R})=\frac{5}{2}\frac{Ze}{R^7}O_{ijk}\int_0^R d^3r\rho({\bf r})o_{ijk}.
\end{equation}
In the above equation the nuclear octupole moment tensor is
\begin{equation}
o_{ijk}=\left[r_i r_j r_k-\frac{r^2}{5}(\delta_{ij}r_k+\delta_{jk}r_i+\delta_{ki}r_j)\right]
\end{equation}
 Taking the screening term into account we can present the screened nuclear octupole moment tensor in the following form:
\begin{equation}
\tilde o_{ijk}=o_{ijk}-\frac{1}{3}\left\{\langle r_i\rangle q_{jk}+\langle r_k\rangle q_{ij}+\langle r_j\rangle q_{ki}\right\}.
\end{equation}
We see that the octupole screening is expressed in terms of the nuclear electric dipole moment
(${\bf d}=Ze\langle{\bf r}\rangle$) and the nuclear quadrupole moment operator $q_{ij}$.
The partially screened octupole potential is given by
\begin{equation}
\varphi^{(3)}({\bf R})=\frac{5}{2}\frac{Ze}{R^7}O_{ijk}\int_0^R d^3r\rho({\bf r})\tilde o_{ijk}.
\end{equation}

The inner part of the nucleus does not give significant contribution to the electron matrix elements of the nuclear octupole field. The situatution is different for the Schiff moment field which was considered in Refs.  \cite{Flambaum:2002,Dmitriev:2005,Flambaum:2012}. To make the picture
for the electrostatic $T,P$-odd nuclear potential complete, we will present a brief derivation for tthe Schiff moment field including the finite nuclear size corrections. The Schiff moment field
is given by the first term  in Eq. (\ref{FinPot})), 
\begin{equation}\label{phi1}
\varphi^{(1S)}({\bf R})\!\equiv\!Ze R_i\!\int_R^{\infty}\!\left(\frac{\langle r_i\rangle}{R^3}-\frac{r_i}{R^3}+\frac{r_i}{r^3}+\frac{\langle r_j\rangle q_{ij}}{r^5}\right)\!\rho({\bf r})d^3r .
\end{equation} 
We see that $\varphi^{(1S)}({\bf R})=0$ if $R>R_N$ (nuclear radius) since $\rho({\bf R})=0$ in that region, i.e. this potential is localized inside the nucleus. 

The electric field of the nuclear Schiff moment polarizes the  atom  and produces atomic EDM.
All  electron orbitals for $l>1$ are extremely small inside the nucleus. Therefore, we should  consider only the matrix elements between $s$ and $p$ Dirac orbitals. We will use the following notations for the electron wavefunctions:
\begin{equation}
\psi({\bf R}) =\begin{pmatrix}
f(R)\Omega_{jlm}\\
-i({\boldsymbol \sigma}\cdot {\bf n})g(R)\Omega_{jlm}
\end{pmatrix}
\end{equation}
where $\Omega_{jlm}$ is a spherical spinor, ${\bf n=R}/R$, $f(R)$ and $g(R)$ are the radial functions. Using  $({\boldsymbol \sigma}\cdot{\bf n})^2=1$ we can write the electron transition density as
\begin{align}\label{Usp}
& \rho_{sp}({\bf R})=\psi_s^{\dagger}\psi_p=\Omega_{s}^{\dagger}\Omega_{p}U_{sp}(R)\\
U_{sp}(R)= & f_s(R)f_p(R)+g_s(R)g_p(R)=\sum_{k=1}^{\infty}b_kR^k
\end{align}
The expansion coefficients $b_k$ can be calculated analytically \cite{Flambaum:2002};  the summation is carried over  odd powers of $k$. Using Eqs. (\ref{phi1},\ref{Usp}) we can find the matrix elements of the electron-nucleus interaction,
\begin{multline}\label{almostlast}
\langle s|-e\varphi^{(1S)}({\bf R})|p\rangle =-Ze^2\langle s|{\bf n}|p\rangle\cdot
\left\{\int_{0}^{\infty}\left[\phantom{\int}\!\!\!\!\!\left(\langle {\bf r}\rangle-{\bf r}\right)\cdot\right.\right.\\
\left.\left.\int_{0}^{r}U_{sp}dR+\left(\frac{{\bf r}}{r^3}+\frac{\langle r_i\rangle q_{ij}}{r^5}\right)\int_{0}^{r}U_{sp}R^3dR\right]\rho d^3r\right\}=\\
-Ze^2\langle s|{\bf n}|p\rangle\cdot\left\{\sum_{k=1}^{\infty}\frac{b_k}{k+1}\left[\langle{\bf r}\rangle\langle r^{k+1}\rangle-\frac{3}{k+4}\langle{\bf r}r^{k+1}\rangle\right.\right.\\
\left.\left.+\frac{k+1}{k+4}\langle r_i\rangle\langle q_{ij}r^{k-1}\rangle\right]\right\},
\end{multline}
where $\langle s|{\bf n}|p\rangle=\int\Omega_s^{\dagger}{\bf n}\Omega_pd\phi\sin\theta d\theta$, $\langle r^n\rangle=\int\rho({\bf r})r^nd^3r$. Note, that all vector values $\langle {\bf r}r^n\rangle$ are due to $P,T$-odd correction $\delta \rho$ to the nuclear charge density $\rho_0$, while $\langle r^n\rangle$ are the usual $P,T$-even moments of the charge density starting from the mean-square radius $\langle r^2\rangle=r_q^2$ for $k=1$.\\

\indent In the limit of the point-like nucleus the Schiff moment potential and its matrix element are given by \cite{Sushkov:1984}:
\begin{align}\label{grad}
\varphi_S({\bf R})&=4\pi{\bf S}\cdot{\boldsymbol\nabla}\delta({\bf R})\\
\langle s|-e\varphi_S|p\rangle&=4\pi e {\bf S}\cdot({\boldsymbol\nabla}\psi_s^{\dagger}\psi_p)_{R=0}.
\end{align}
 For the solutions of the  Dirac equation the product $({\boldsymbol\nabla}\psi_s^{\dagger}\psi_p)_{R\rightarrow0}$ is infinite for a point-like nucleus. Therefore,  we need a finite-size Schiff moment potential. We have  shown \cite {Flambaum:2002} that this potential increases linearly inside the nucleus and vanishes at the nuclear surface. We suggested an approximate expression for such potential which is convenient for the calculations of atomic EDM:  
 \begin{equation}\label{Finite1}
\varphi_S({\bf R})=-\frac{3{\bf S'}\cdot {\bf R}}{B}n(R),
\end{equation}
where $B=\int n(R)R^4dR\approx R_N^5/5$, $R_N$ is the nuclear radius and $n(R)$ is a smooth function equal to 1 for $R<R_{N}-\delta$ and 0 for $R>R_{N}+\delta$; $n(R)$ can be taken as proportional to the nuclear density $\rho_0$ (note that we can choose any normalization of $n(r)$ since the normalization constant cancels out in the ratio $n/B$, see Eq. (\ref{Finite1})). 

\FloatBarrier
\begin{table*}[tb]\center
\caption{Schiff moment $S$ and the ratio of the $Z\alpha$  correction $\triangle S=S'-S$  to $S$ for  $^{205}$Tl and  $^{199}$Hg nuclei. $S_0$ and $\triangle S_0$ are the bare values, without the core polarization correction produced by the strong residual nuclear forces. $S_{tot}$ and $\triangle S_{tot}$ are the  results including the core polarization effects.  In brackets there are  values of $L'/S$ ( where $L'=L-S$) which are the results of the calculations from \cite{Dmitriev:2005}. To obtain the final values of $S$ and $S'=S+ \triangle S$ one should sum the contributions of the three interaction constants $g_0$, $g_1$ and $g_2$. The units of $S$ are $e\cdot fm^3$.}
{\renewcommand{\arraystretch}{0}%
\begin{tabular}{l l r r l r r l}
\hline\hline
\rule{0pt}{4pt}\\
& & $\quad\qquad S_0$ & $\quad\triangle S/S_0$&$(L'/S_0)$ & $\qquad S_{tot}$ & $\quad\triangle S_{tot}/S_{tot}$&$(L'_{tot}/S_{tot})$\\
\rule{0pt}{4pt}\\
\hline
\rule{0pt}{4pt}\\
& $g_0$ & -0.085 & 0.05&(-0.1) & -0.006 & 0.09&(-0.05)\\ 
\rule{0pt}{2pt}\\
$^{199}$Hg$\qquad$ & $g_1$ & -0.085 & 0.05&(-0.1) & -0.036 & 0&(-0.15)\\
\rule{0pt}{2pt}\\
& $g_2$ & 0.17 & 0.05&(-0.1) & 0.019 & 0.06&(-0.08)\\
\rule{0pt}{4pt}\\
\hline
\rule{0pt}{4pt}\\
& $g_0$ & -0.075 & 0.05&(-0.09) & -0.014 & -0.03&(-0.18)\\ 
\rule{0pt}{2pt}\\
$^{205}$Tl & $g_1$ & -0.028 & 0.23&(-0.39) & -0.082 & -0.03&(-0.18)\\
\rule{0pt}{2pt}\\
& $g_2$ & 0.237 & 0.06&(-0.08) & -0.007 & -0.34&(-0.51)\\
\rule{0pt}{4pt}\\
\hline\hline
\end{tabular}}\label{Tab:1}
\end{table*}

\indent We now set the matrix elements (\ref{almostlast}) of the true nuclear $T,P$-odd potential (\ref{phi1}) to be equal to the matrix elements of the effective potential   (\ref{Finite1}) which are given by 
\begin{equation}\label{last}
\begin{split}
\langle s|-e\phi({\bf R})|p\rangle&=15e\langle s|{\bf n}|p\rangle\cdot\frac{{\bf S'}}{R_N^5}\int_0^{\infty}U_{sp}R^3n(R)dR\\
&=15e\langle s|{\bf n}|p\rangle\cdot{\bf S'}\sum_{k=1}^{\infty}b_k\frac{R_N^{k-1}}{k+4},
\end{split}
\end{equation}
where we have made approximation $\int n(R)R^kdR\approx R_N^{k+1}/(k+4)$. Equating (\ref{almostlast}) and (\ref{last}) we obtain
\begin{equation}\label{Sprime}
\begin{split}
&{\bf S'}=\frac{Ze}{15}\frac{1}{\sum_{k=1}^{\infty}\frac{b_k}{b_1}\frac{1}{k+4}R_N^{k-1}}\sum_{k=1}^{\infty}\frac{b_k}{b_1}\frac{1}{k+1}\\
&\left[\frac{3}{k+4}\langle{\bf r}r^{k+1}\rangle-\langle{\bf r}\rangle\langle r^{k+1}\rangle-\frac{k+1}{k+4}\langle r_i\rangle\langle q_{ij}r^{k-1}\rangle\right]
\end{split}
\end{equation}
Note that $S'$ in Eq.  (\ref{Sprime}) differs from the local dipole moment $L$ defined in Ref. \cite{Flambaum:2002} and calculated in \cite{Dmitriev:2005}. The  local dipole moment $L$ does not contain the sum in the denominator in  Eq.  (\ref{Sprime}) and corresponds to the 
$\delta$-function form (similar to Eq. (\ref{grad})) of the effective Schiff moment potential.

In light atoms  $(Z\alpha \ll 1)$ it is sufficient to keep  $b_1$ only.  This gives us the well-known expression for the Schiff moment  \cite{Sushkov:1984}:
\begin{equation}
{\bf S_0}\equiv\frac{Ze}{10}\left[\langle{\bf r}r^2\rangle-
\frac{5}{3}\langle{\bf r}\rangle\langle r^2\rangle-\frac{2}{3}\langle r_i\rangle\langle q_{ij}\rangle\right].
\end{equation}
The first  correction is given by  the ratio $b_3/b_1$. This ratio is different for matrix elements $s$ - $p_{1/2}$ 
($b_3/b_1=-(3/5) Z^2\alpha^2/R_N^2$)  and  $s$ - $p_{3/2}$ ($b_3/b_1=-(9/20) Z^2\alpha^2/ R_N^2$).  However, with the 10\% accuracy we can use the  average of these two values $b_3/b_1\approx -0.5 Z^2\alpha^2/R_N^2$. This gives 
\begin{eqnarray}\label{Schiff1}
\nonumber
{\bf S'}=\frac{Ze}{10}\frac{1}{1-\frac{5}{14}Z^2\alpha^2}
\left\{\left[\langle{\bf r}r^{2}\rangle-\frac{5}{
3}\langle{\bf r}\rangle\langle r^{2}\rangle-\frac{2}{3}\langle r_i\rangle\langle q_{ij}\rangle\right]\right.\\
\nonumber
\left.-\frac{5}{28}\frac{Z^2\alpha^2}{R_N^2}\left[\langle{\bf r}r^{4}\rangle-\frac{7}{3
}\langle{\bf r}\rangle\langle r^{4}\rangle-\frac{4}{3}\langle r_i\rangle\langle q_{ij}r^2\rangle\right]\right\}\\
\end{eqnarray}
This equation allows one to calculate the  $Z\alpha$ corrections to the Schiff moment. 
In our work \cite{Dmitriev:2005} such calculations were performed for the local dipole moment $L$.
Corresponding expression for the local dipole moment $L$ \cite{Flambaum:2002,Dmitriev:2005} does not contain the factor  $(1-5Z^2\alpha^2/14)$ in the denominator (see Eq. (\ref{Schiff1}) for $S'$). All recent atomic EDM calculations  \cite{Dzuba:2009,Dzuba:2007,Dzuba:2002} have been performed using the Schiff moment potential in the form (\ref{Finite1}).
Therefore, we calculate the values of $S'$  for two nuclei of experimental interest, $^{199}$Hg (with valence neutron) and $^{205}$Tl (with valence proton), using  calculations of $L$ in Ref. \cite{Dmitriev:2005} .

In \cite{Dmitriev:2005}  we used a finite-range $P,T$-violating nucleon-nucleon interaction of the form
\begin{equation}
\begin{split}
W&({\bf r}_a-{\bf r}_b)=-\frac{g_s}{8\pi m_p}\left\{\left[g_0{\bf \tau_a\cdot\tau_b}+g_2\left({\bf \tau_a\cdot\tau_b}-3\tau_a^z\tau_b^z\right)\right]\right.\\
&\left.\times({\boldsymbol\sigma}_a-{\boldsymbol\sigma}_b)+g_1\left(\tau_a^z{\boldsymbol\sigma}_a
-\tau_b^z{\boldsymbol\sigma}_b\right)\right\}\cdot{\boldsymbol\nabla}_a\frac{e^{-m_{\pi}r_{ab}}}{r_{ab}}\,,
\end{split}
\end{equation}
where $m_p$ is the proton mass and $r_{ab}=|{\bf r}_a-{\bf r}_b|$. The core polarization corrections
produced by the strong residual nuclear forces have been calculated using the RPA technique.

Results of the calculations of the  Schiff moment and $Z\alpha$ corrections are presented in Table \ref{Tab:1}. As one can see, in most cases use of $S'$ (instead of $L$) leads to smaller  values of the $Z\alpha$ corrections. For $S'$ typical values of the  $Z\alpha$ corrections are  about 5-10\%.
Larger $Z\alpha$ corrections (up to 34\% for $S'$ and 51\% for $L$) appear in the cases where the main contributions are suppressed.

\end{document}